# Electro-optic switching of dielectrically negative nematic through nanosecond electric modification of order parameter

Bing-Xiang Li[1,2], Volodymyr Borshch[1], Sergij V. Shiyanovskii[1], Shao-Bin Liu[2], and Oleg D. Lavrentovich[1,*]

[1]Chemical Physics Interdisciplinary Program, Liquid Crystal Institute, Kent State University, Kent, Ohio 44242, USA

[2]College of Electronic and Information Engineering, Nanjing University of Aeronautics and Astronautics, Nanjing 210016, China

We present experimental studies of nanosecond electric modification of the order parameter (NEMOP) in a variety of nematic materials with negative dielectric anisotropy. The study demonstrates that NEMOP enables a large amplitude of fast (nanoseconds) electro-optic response with the field-induced birefringence on the order of 0.01 and a figure of merit (FoM) on the order of $10^4 \ \mu m^2/s$; the latter is orders of magnitude higher than the FoM of the Frederiks effect traditionally used in electro-optic nematic devices. The amplitude of the NEMOP response is generally stronger in nematics with larger dielectric anisotropy and with higher natural (field-free) birefringence.

Nematic liquid crystals (NLCs) have numerous electro-optic applications enabled by anisotropy of their properties[1]. The optic axis of the NLC is called the director $\hat{\mathbf{n}}$. In typical electro-optic applications, an AC electric field applied to a dielectrically anisotropic NLC is used to realign the director and thus to change the effective birefringence $\delta n$ or phase retardance $\Gamma = d\delta n$, where $d$ is the length of pathway of light in the NLC[1]. Very often, in applications such as displays, optical shutters, modulators, switches, and beam steerers, a desirable mode of operation is to switch large retardance within a short period of time, characterized by a figure of merit[1] FoM = $\Gamma^2 / (\pi^2 \tau_{off})$, where $\tau_{off}$ is the relaxation time of retardance to its field-free state. The process of dielectric reorientation is relatively slow, on the scale of milliseconds, especially during the field-off stage. The typical FoM is on the order of a few $\mu m^2/s$.[1] The speed of switching can be accelerated by a variety of approaches, such as optimizing the viscoelastic parameters of the NLCs,[1] overdriving,[1] realigning an NLC in a submicrometer-templated polymer network[2] with the response time on the order of 0.1 ms or by employing a dual-frequency NLC in a special geometry with a high pretilt angle in which case the FoM can reach $10^3$ $\mu m^2/s$.[3] Recently, a different approach to change the optical retardance of the NLC has been proposed[4], based on electric modification of the order parameter (EMOP)[5-11]. In this approach, the electric field does not change the director orientation but modifies the optic tensor[11], with a very fast (tens of nanoseconds) response[4]. So far, this nanosecond EMOP (NEMOP) effect has been demonstrated only for one material[4], CCN-47. As compared to the regular dielectric reorientation (Frederiks effect), NEMOP response in this material is much faster (nanoseconds instead of milliseconds), but requires higher electric fields ($E = 10^8$ V/m instead of $10^6$ V/m) and produces lower effective birefringence ($\delta n$ = 0.001 instead of 0.1)[4]. The goal of this work is to explore whether the amplitude of NEMOP effect can be enhanced by a targeted choice of nematic materials. With this goal in mind, we present the experimental data on a large number of materials that differ in the magnitude of negative dielectric anisotropy $\Delta\varepsilon$ and natural (field-free) birefringence $\Delta n$. The study reveals that strongly anisotropic materials allow one to achieve a high $\delta n$ = 0.01, with FoM raising to $10^4$ $\mu m^2/s$. The achieved field-induced birefringence demonstrates that the NEMOP effect can be effectively used in practical applications in which the level of switchable retardance is on the order of half-wavelength of light.

---

* Electronic mail: olavrent@kent.edu

In the experiments, we use planar cells of NLC with negative dielectric anisotropy $\Delta\varepsilon = \varepsilon_{//} - \varepsilon_{\perp}$; the subscripts refer to the field orientation parallel and perpendicular to the director $\hat{\mathbf{n}}$, respectively. The cells are comprised of two glass plates with transparent indium tin oxide (ITO) electrodes of low resistivity (between 10 and 50 Ω/sq). The area of electrodes is $A = 2 \times 2$ mm$^2$. The inner surfaces of the cell are coated with layers of unidirectionally rubbed polyimide PI-2555 (HD MicroSystems). The cells are assembled in a parallel fashion. The field applied across the cell does not change the orientation of $\hat{\mathbf{n}}$ and only modifies the orientational order.

We explore materials of different birefringence $\Delta n = n_{//} - n_{\perp}$ and dielectric anisotropy $\Delta\varepsilon$, as specified in Table I: HNG705800-100, HNG715600-100 (purchased from Jiangsu Hecheng Display Technology), MCT-5 (Kingston Chemicals), MLC-2079, MLC-2080, ZLI-2806, ZLI-4330, MJ961200, MJ97731, MJ951152, MAT-03-382, MAT-08-192 (all Merck), and CCN-47 (Nematel GmbH). The RC time $\tau_{RC}$ was less than 4 ns for all cells.

In order to measure the electric field induced optic response, we use a laser beam (He-Ne, $\lambda = 632.8$ nm) linearly polarized along the direction that makes an angle of 45° with the incidence plane of the NLC slab, Fig.1. The beam passes through the cell, Soleil-Babinet compensator, and the analyzer crossed with the polarizer, see Ref. 4 for the detailed description. The transmitted light intensity is measured using a photodetector TIA-525 (Terahertz Technologies, response time < 1 ns). The test cells are sandwiched between two prisms, so that the incident angle of light beam is 45° with respect to the normal to the cell, in order to eliminate the contribution of director fluctuations to the optical response[4], Fig. 1.

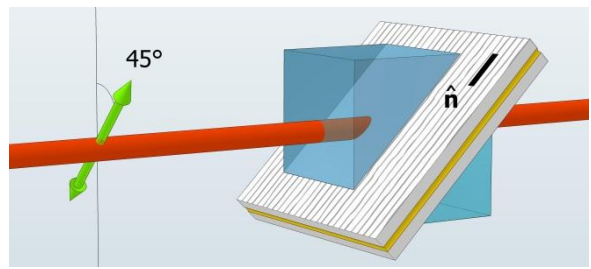

FIG. 1. Experimental setup: a test cell sandwiched between two right angle prisms, probed with a linearly polarized laser beam that propagates inside the nematic slab at the angle 45 ° with respect to the cell normal.

The temperature of cell assemblies is controlled with Linkam LTS350 hot stage. A voltage pulse of duration 400 ns is applied using a pulse generator HV 1000 (Direct Energy). The pulse generator creates voltage pulses with sharp rise and fall edges of characteristic time 1 ns. The applied voltage pulses and photodetector signals were measured with 1G samples/s digital oscilloscope TDS2014 (Tektronix).

The NEMOP response is observed in all the materials studied, Table I and Fig. 2, 3. For a given driving voltage applied to cells of approximately the same thickness (ranging from 4.5 to 5.1 μm), the field-induced birefringence $\delta n$ depends strongly on the NLC used. Among all the studied materials, the mixture HNG715600-100 displays the largest NEMOP effect, with $\delta n \approx 0.013$ at applied voltage pulse of amplitude $U_0 = 873$ V, which is an order of magnitude higher than the one achieved in CCN-47, Fig. 2(a) and Table I.

Fig. 2(b) shows the optical response of HNG715600-100 cell of thickness 5.1 μm to the applied voltages of different amplitude $U_0$. The experimentally determined profile of the optical response is well fitted by the theoretical model[4], in which

$$\delta n(t) = \frac{1}{\sqrt{2}\bar{n}_{LC}} \left( \frac{1}{2} \delta\tilde{\varepsilon}_u(t) + \frac{3}{4} \delta\tilde{\varepsilon}_b(t) \right), \quad (1)$$

where $\bar{n}_{LC}$ is the average refractive index of NLC, $\delta\tilde{\varepsilon}_u$ and $\delta\tilde{\varepsilon}_b$ are the uniaxial and biaxial contributions to the optic tensor. The dynamics of $\delta\tilde{\varepsilon}_u$ and $\delta\tilde{\varepsilon}_b$ is modeled within the Landau-Khalatnikov approach[12]

$$\tau_i \frac{d\delta\tilde{\varepsilon}_i(t)}{dt} = \alpha_i E^2(t) - \delta\tilde{\varepsilon}_i(t), \quad (2)$$

with the solution

$$\delta\tilde{\varepsilon}_i(t) = \int_0^t \frac{\alpha_i E^2(t')}{\tau_i} \exp\left(\frac{t'-t}{\tau_i}\right) dt'. \quad (3)$$

Here the subscript "$i$" reads either "$u$" or "$b$", depending on the uniaxial or biaxial nature of the contribution; $\tau_u$ and $\tau_b$ are the corresponding relaxation times; $\alpha_u$ and $\alpha_b$ are the corresponding susceptibilities to the applied field[4]. Note that in the model of NEMOP effect, the on and off times are equal to each other, which is supported by the experimental data for the materials studied in this work. With the known $E(t)$, the experimental $\delta n(t)$ is fitted using Eqs. (1, 3) with the fitting parameters $\tau_u$, $\tau_b$, $\alpha_u$, and $\alpha_b$. An example of such a fitting is presented in Fig. 2(b), for an optical response of HNG715600-100 to the voltage pulse of amplitude 873 V and duration 400 ns. In this case, $\tau_u = 33\,\text{ns}$, $\tau_b = 3\,\text{ns}$, $\alpha_u = 9.2 \times 10^{-19}\,\text{m}^2/\text{V}^2$, and $\alpha_b = 6.8 \times 10^{-19}\,\text{m}^2/\text{V}^2$. Note that for some materials, $\tau_b$ might be 1 ns or even smaller, but the limitations of our experimental set-up do not allow us to resolve time scales shorter than 1 ns. The close fitting seen in Fig.1b demonstrates that the model Eq. (1-3) captures the essential features of the physical mechanisms very well, despite the fact that most of the materials in Table I represent complex mixtures of different chemicals.

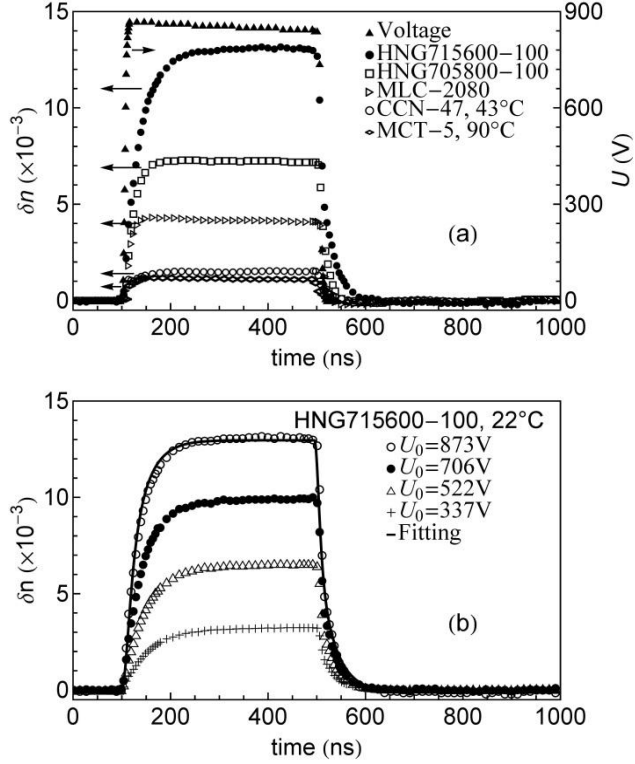

FIG. 2. (a) Dynamics of field-induced birefringence $\delta n(t)$ in response to a voltage pulse $U(t)$ (filled triangles) for HNG715600-100 (disks), HNG705800-100 (open squares), MLC-2080 (open triangles), CCN-47 (open circles), and MCT-5 (diamonds). (b) Dynamics of $\delta n(t)$ for HNG715600-100 in response to voltage pulses of amplitudes $U_0 = 337$V (crosses), 522V (open triangles), 706V (disks), and 873V (open circles). Fitting with $\tau_u = 33$ ns, $\tau_b = 3$ ns, $\alpha_u = 9.2 \times 10^{-19}$ m$^2$/V$^2$, and $\alpha_b = 6.8 \times 10^{-19}$ m$^2$/V$^2$ is shown by the solid line. All data were determined at room temperature (22 °C), except those for CCN-47 (43 °C) and MCT-5 (90 °C).

The two important parameters of the NEMOP response are the amplitude of field-induced birefringence $\delta n$ and the response time $\tau$, Fig. 3. The voltage dependence of $\delta n$ is generally quadratic, Fig. 3(a), with some deviations observed at high voltages. The highest values of $\delta n$ are achieved in the two mixtures HNG715600-100 and HNG05800-100 which also demonstrate the highest dielectric anisotropy. However, the dielectric anisotropy is not the only factor that determines the amplitude of NEMOP: As seen from Table I, CCN-47 with high dielectric anisotropy yields a relatively weak response. The natural birefringence $\Delta n$ is also not a clear single indicator of how strong is the NEMOP response might be, as HNG705800-100 with a relatively modest $\Delta n = 0.08$ produces a high $\delta n \approx 0.008$.

The dynamics of optical response is determined by the uniaxial and biaxial modifications of the order parameter, so that $\tau = \max(\tau_u, \tau_b)$. In most of the studied materials, the slowest process is the relaxation of uniaxial contribution with the characteristic time $\tau_u$ that is on the order of tens of nanoseconds, Fig. 3(b), much slower than $\tau_b \sim 1\,\text{ns}$. As seen in Fig. 3(b), the values of $\tau$ are voltage-independent, as expected by the model. In some materials, such as MAT-08-192, $\tau$ is extremely short, 1 ns or so, which is again an order of magnitude better than the case of the previously studied CCN-47, Table I. Extreme smallness of $\tau$ requires further exploration of the physical mechanisms behind the order parameter dynamics in these materials.

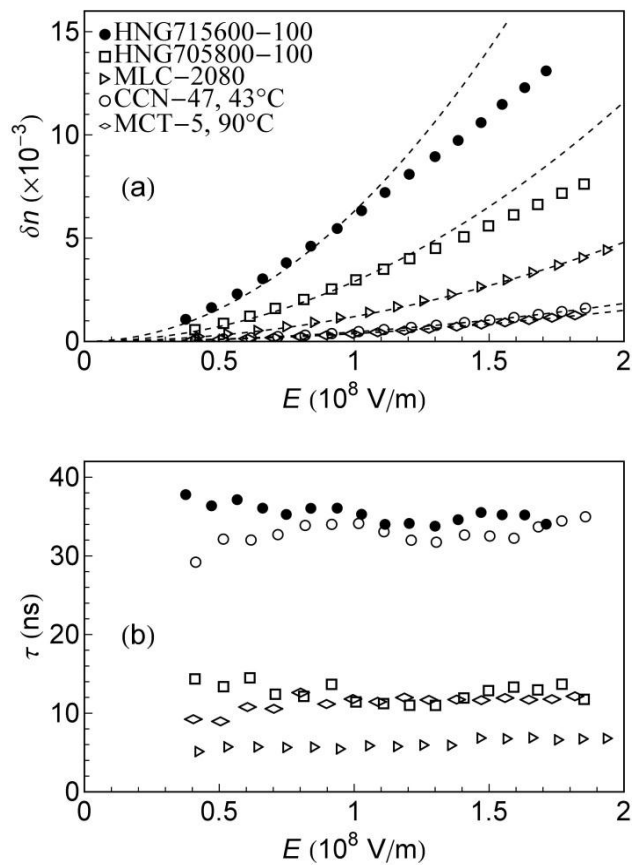

FIG. 3. Electric field dependence of (a) the field-induced birefringence $\delta n$ and (b) response time $\tau$ for HNG715600-100 (disks), HNG705800-100 (open squares), MLC-2080 (open triangles), CCN-47 (open circles), and MCT-5 (diamonds). Experimental data were measured at the room temperature (22 ℃), except those for CCN-47 (43 ℃) and MCT-5 (90 ℃). The dashed lines in (b) are the corresponding parabolas.

TABLE. I. NEMOP parameters and material characteristics of NLCs.

| Material [a] | $E$ ($10^8$ V/m) | $\delta n$ ($10^{-3}$) | $\tau$ (ns) | FoM ($10^3$ μm$^2$/s) | $\Delta\varepsilon$ @ 1 kHz | $\Delta n$ @ 589 nm |
|---|---|---|---|---|---|---|
| HNG715600-100 | 1.7 | 13.2 | 33 | 27 | -12.2 | 0.15 |
| HNG705800-100 | 1.8 | 7.6 | 12 | 20 | -9.2 | 0.08 |
| MLC-2080 | 1.9 | 4.5 | 6 | 14 | -6.4 | 0.11 |
| MLC-2079 | 1.9 | 4.2 | 5 | 15 | -6.1 | 0.15 |
| MJ961200 | 1.9 | 3.2 | 11 | 3.8 | -5.6 | 0.11 |
| MJ951152 | 1.9 | 2.8 | 2.5 | 12 | -4.2 | 0.08 |
| MAT-08-192 | 1.9 | 2.5 | 1 | 25 | -3.9 | 0.17 |
| MAT-03-382 | 1.9 | 2.1 | 2 | 8.9 | -3.8 | 0.08 |
| MJ97731 | 1.3 | 1.7 | 10 | 2.4 | -5.0 | 0.10 |
| CCN-47 | 1.9 | 1.5 | 34 | 0.3 | -5.1 | 0.03 |
| MCT-5 | 1.8 | 1.3 | 12 | 0.6 | -2.5 | 0.20 |
| ZLI-4330 | 1.9 | 1.0 | 3 | 1.3 | -1.9 | 0.15 |
| ZLI-2806 | 1.2 | 0.3 | 2 | 0.2 | -4.8 | 0.04 |

[a] at 20 ˚C, except for CCN-47 (40 ˚C) and MCT-5 (90 ˚C).

To conclude, we explored the nanosecond switching in a number of NLCs with negative dielectric anisotropy in which the applied electric field causes uniaxial and biaxial modifications of the order parameter but does not realign the director. We found that all of the explored nematics demonstrate a substantial field-induced birefringence, ranging from 0.01 to 0.001 at applied field on the order of ~$10^8$V/m. The corresponding FoM that characterizes how much of optical retardance can be switched within a certain time, is on the order of $10^4$ μm$^2$/s, which is at least one order of magnitude higher than the FoM for dual-frequency nematics[3] and two-three orders of magnitude higher than the values achieved in Frederiks switching of regular nematics[1]. The significant improvement of the FoM is rooted in the very nature of the NEMOP effect in which the dynamics is controlled by the molecular response and in which the field-on and field-off relaxation times are essentially the same, being in the range of nanoseconds and tens of nanoseconds. The latter fact is a distinct beneficial feature of the NEMOP electro-optics, since in the Frederiks effects involving director reorientation within the timescale of milliseconds, the field-on time can be accelerated by a very high field, but the field-off state is typically much slower.[1] Instead of the voltage pulses, one could also employ optical light pulses. The efficiency and versatility of optical driving can be greatly expanded by using photosensitive azbenzene-based LCs that respond to nanosecond laser pulses and show a broad variety of relaxation times ranging from 1 ms to many hours, depending on the chemical structure, as described by Hrozhyk et al [13].

The very high FoM demonstrated in this work is achieved in relatively thin cells of thickness about 5 μm. Note that the switching time in the NEMOP effect does not depend on the thickness of the cell, which allows one to further increase the FoM defined in the case of NEMOP as $(d\delta n)^2 / \pi^2 \tau$, by increasing the pathway of light $d$ (say, by using thicker cells) while preserving the time of switching $\tau$. In the regular Frederiks effect, the response time typically grows as the square of the cell thickness.

The data also demonstrate that the electro-optic performance of NEMOP effects depends strongly on the material parameters such as dielectric and optical anisotropy. The level of optimization achieved in this work might be potentially advanced by further exploration of different materials. The NEMOP effect can enable ultrafast electro-optic effects in applications ranging from displays to shutters, limiters, modulators, switches, and beam steerers, as the switchable optical retardance reaches the required levels of half-wavelength.

**Acknowledgements.** The work was supported by DOE grant DE-FG02-06ER 46331, State of Ohio through Ohio Development Services Agency and Ohio Third Frontier grant TECG20140133, and by China Scholarship Council and Jiangsu Innovation Program for Graduate Education, grant No.CXZZ13_0155. The content of this publication reflects the views of the authors and does not purport to reflect the views of Ohio Development Services Agency and/or that of the State of Ohio.